\documentclass[proof]{WileyASNA-v1}

\articletype{ORIGINAL ARTICLE}%

\received{xx xxxx xx}
\revised{xx xxxx xx}
\accepted{xx xxxx xx}

\begin{document}

\title{Discovery of eclipses in the cataclysmic variable LAMOST J035913.61+405035.0}

\author{V. P. Kozhevnikov}

\authormark{KOZHEVNIKOV\textsc{}}

\address{\orgdiv{Astronomical Observatory}, \orgname{Ural Federal University}, \orgaddress{\state{Yekaterinburg}, \country{Russia}}}

\corres{V. P. Kozhevnikov, Ural Federal University, 19 Mira Street, 620002  Yekaterinburg, Russia \email{valery.kozhevnikov@urfu.ru}}


\abstract{I conducted photometric observations of the cataclysmic variable LAMOST J035913.61+405035.0 and discovered previously unknown eclipses. During these observations, I recorded 14 eclipses over two groups of nights separated by 13 months. I accurately determined the orbital period of the system to be $P_{\rm orb}=0.228\,343\,85\pm 0.000\,000\,21$~d. For the eclipses, I derived an ephemeris which is valid for a long time and suitable for studying changes in the orbital period. The out-of-eclipse magnitude of the star varied between 15.32$\pm$0.02 and 17.25$\pm$0.08~mag. As the brightness decreased, the eclipses became deeper and narrower. The average depth of eclipses was 1.35$\pm$0.10~mag, and the average width at half-depth was 16.9$\pm$0.7~min. I estimated the range of possible orbital inclinations to be between $72.8^\circ$ and $76.0^\circ$, and the range of average absolute $V$-band magnitudes of the disc to be between 5.16$\pm$0.15 and 5.44$\pm$0.15~mag. Although based on the light curve from the ZTF survey, LAMOST J035913.61+405035.0 showed only small outbursts with amplitudes below 1.5~mag, it should be classified as a dwarf nova because the average disc brightness and mass transfer rate were below the limit of thermal instability. However, there have been no significant outbursts in the ZTF light curve over the past 1.6~yr. Instead, a gradual decrease in brightness lasting 120~d suggests that this object may occasionally become a nova-like variable of the VY~Scl type.}

\keywords{Stars: individual, LAMOST J035913.61+405035.0; Novae, Cataclysmic variables; Binaries: eclipsing }

\jnlcitation{\cname{%
\author{Kozhevnikov V. P.}, 
} (\cyear{2024}), 
\ctitle{Discovery of eclipses in the cataclysmic variable LAMOST J035913.61+405035.0}, \cjournal{Astron. Nachr.}, \cvol{xxxx;xx:x-x}.}


\maketitle


\section{Introduction}\label{sec1}

Cataclysmic variables (CVs) are binary star systems consisting of a white dwarf and a red dwarf. The red dwarf fills its Roche lobe and transfers matter to the white dwarf, forming a bright disc around the white dwarf. If the white dwarf has a strong magnetic field, this disc may be truncated or replaced by a stream, as seen in polars and intermediate polars. Due to non-synchronous rotation, some intermediate polars may exhibit coherent short-period oscillations related to the spin of the white dwarf. All CVs show rapid, non-periodic brightness changes called flickering. CVs can be classified as dwarf novae or nova-like variables, depending on their photometric behaviour. Dwarf novae show outbursts lasting from 2 to 20~d with amplitudes ranging from 2--5 mag. Nova-like variables may not exhibit outbursts or, if they do, their amplitude is less than 1 mag \citep{honeycutt01}. However, some nova-like variables, such as VY~Scl stars, may experience occasional brightness decreases, dropping more than 1~mag from their maximum brightness. Classical novae after an eruption behave similarly to nova-like variables. When the orbital inclination of a CV is large, eclipses may occur. Eclipsing CVs provide valuable information about the orbital period and orbital inclination. Detailed discussions of CVs can be found in \citet{ladous94}, \citet{warner95}, \citet{hellier01} and more recently \citet{hardy17}, which presents recent data on eclipsing CVs.


\citet{hou20} identified 58 new CV candidates using the Large Sky Area Multi-Object Fibre Spectroscopic Telescope (LAMOST) survey. One of these candidates, LAMOST J035913.61+405035.0 (Gaia DR3 229960043749437952), hereafter J0359, shows a noticeable HeII emission line at 4686\AA, suggesting a possible magnetic nature. Its coordinates are suitable for observations from our observatory, which is located 80 km from Ekaterinburg. I conducted photometric observations of J0359 to search for short-period oscillations, similar to those seen in intermediate polars. Although the LAMOST survey did not detect any double-peaked spectral lines indicative of a high orbital inclination for this object, my photometry revealed previously unknown eclipses. To accurately determine the eclipse period, I continued my observations over 11 nights for 13 months. This paper presents the results of these observations.


\section{Observations}\label{sec2}

For observations of variable stars, I use a three-channel photometer equipped with photomultipliers. This device works by converting photons into electrical pulses, which allows for accurate counting of photons and measurements of light intensity. The design and noise analysis of the photometer are described in \citet{kozhevnikoviz00}. The photometer continuously measures the brightness of two stars and the sky background in the field of view of the telescope. This technique provides high precision even under challenging conditions, such as unstable atmospheric transparency and a changeable sky background. Three pulse counters send data to a computer. The photometer works with a 70-cm Cassegrain telescope, which is equipped with computer-controlled step motors. A CCD guiding system is used to correct tracking errors and accurately centre two stars in the photometer's diaphragms during observations. All components, including the telescope, photometer and guiding system, operate automatically under computer control. Human intervention is required only if thick clouds interfere with the observation process.

I performed photometric observations using white light without filters (approximately 3000--8000\AA), with a time resolution of 16~s. I measured the light intensity of two stars using a 16-arcsec diaphragm for each, and the intensity of the sky background using either a 30- or 40-arcsec diaphragm, depending on the year. These larger diaphragms help to reduce the photon noise caused by Poisson fluctuations in sky background photons. On the first night of observations, J0359 was faint and invisible to the eye (about 17~mag). Nevertheless, the total light intensity of this star and the sky background was noticeably higher than that of the sky background alone, allowing me to locate and centre J0359 within the photometer's diaphragm by moving the telescope according to the stellar coordinates. That night, I observed an eclipse in J0359. Eclipses in J0359 were not previously known, so I continued observing to accurately determine their period. They lasted for a total of 61 hours over 11 nights, covering 13 months, as shown in the observational log in Table~\ref{tab1}.

\begin{table}[t]
\caption{Log of the observations}
\label{tab1}
\begin{tabular}{@{}l c c c}
\hline
\noalign{\smallskip}
Date (UT) & BJD$_{\rm TDB}$ start & Length    & Out-of-ecl. \\ 
                 & (-2459000)                  &   (h)       &   magnitude \\
\noalign{\smallskip}
\hline
\noalign{\smallskip}
2021 Oct. 7     &   495.176991    &   8.2   &     17.03$\pm$0.10   \\
2021 Oct. 8     &   496.168596    &   4.2   &     17.05$\pm$0.04   \\
2021 Oct. 13   &   501.214034    &   3.7  &      16.97$\pm$0.04   \\
2021 Nov. 2    &   521.165145    &   9.7   &      15.43$\pm$0.04  \\
2021 Nov. 3    &   522.256227    &   7.4   &      15.32$\pm$0.02  \\
2022 Sep. 29   &   852.376261    &   2.9  &      16.26$\pm$0.03  \\
2022 Oct. 17    &   870.181786    &   1.7   &     16.55$\pm$0.05  \\
2022 Nov. 16   &   900.134186    &   3.1  &      17.12$\pm$0.03  \\
2022 Nov. 17   &   901.204404    &   8.9  &      17.15$\pm$0.02  \\
2022 Nov. 18   &   902.389911    &   2.3  &      17.15$\pm$0.03  \\
2022 Nov. 19   &   903.079156    &   9.0  &      17.25$\pm$0.08  \\
\noalign{\smallskip}
\hline
\end{tabular} 
\end{table}


\begin{figure}[t]
\centering
\includegraphics[width=84mm]{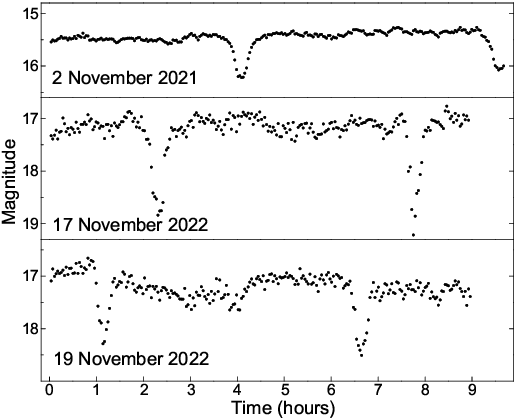}
\caption{J0359 light curves showing pairs of eclipses} 
\label{fig1}
\end{figure}

\section{Analysis and results}\label{sec3}

The photon counts must follow a Poisson distribution, with $N=\sigma^2$. Here $N$ is the average photon count and $\sigma$ is the standard deviation of the data points in the time series. Using a stable light source, I found that $N/\sigma^2$=0.98$\pm$0.01 in my photometer, which allows me to estimate the photon noise. To calculate the photon noise in one of the star channels of the photometer, in magnitudes, I used the following formula: $\sigma_{\rm photon}=1.08 \sqrt {(N_{\rm total}+k^2 N_{\rm sky})} / ( N_{\rm total} - k N_{\rm sky})$,  where $N_{\rm total}$ is the total average count of photons in the star channel, $N_{\rm sky}$ is the average count of sky background photons in the sky background channel and $k$ is a factor that accounts for differences in the sky background between the two channels. This formula was tested using a simulated data set that mimics the actual photon counts from observations.

\begin{table}[t]
\caption[ ]{Parameters of the observed eclipses}
\label{tab2}
\begin{tabular}{@{}l c c c}
\hline
\noalign{\smallskip}
Date (UT)  &   Depth      & Width      &  \scriptsize BJD$_{\rm TDB}$ mid-ecl.   \\
                  &  (mag)      &  (min)     & (-2459000)                        \\
\noalign{\smallskip}
\hline
\noalign{\smallskip}
2021 Oct. 7         & 1.69$\pm$0.06     & 16.1$\pm$0.8   &   495.30445(19)     \\
2021 Oct. 8        & 1.37$\pm$0.05     & 15.1$\pm$0.8    &   496.21805(19)    \\
2021 Oct. 13       & 1.17$\pm$0.05     & 17.1$\pm$1.0    &   501.24118(26)    \\
2021 Nov. 2a      & 0.82$\pm$0.01      & 19.3$\pm$0.5    &   521.33470(10)    \\
2021 Nov. 2b      & 0.75$\pm$0.01      & 21.9$\pm$1.0    &   521.56405(29)    \\
2021 Nov. 3        & 0.76$\pm$0.02      & 22.1$\pm$0.8    &   522.47845(21)    \\
2022 Sep. 29      & 1.67$\pm$0.03      & 16.6$\pm$0.5    &   852.43400(10)    \\
2022 Oct. 17      & 1.52$\pm$0.10      & 17.3$\pm$1.8    &   870.24610(37)    \\
2022 Nov. 16       & 1.36$\pm$0.10    & 15.3$\pm$1.5   &   900.15701(36)    \\
2022 Nov. 17a     & 1.77$\pm$0.08     & 16.8$\pm$1.0   &   901.30034(24)     \\
2022 Nov. 17b     & 1.86$\pm$0.08     & 13.3$\pm$0.8   &   901.52778(19)    \\
2022 Nov. 18       & 1.50$\pm$0.06     & 17.8$\pm$0.8   &   902.44137(22)    \\
2022 Nov. 19a     & 1.36$\pm$0.10     & 13.1$\pm$1.3   &   903.12686(31)   \\
2022 Nov. 19b     & 1.36$\pm$0.06     & 14.8$\pm$0.8   &   903.35527(22)   \\
\noalign{\smallskip}
\hline
\end{tabular}
\end{table}

\begin{figure}[t]
\centering
\includegraphics[width=84mm]{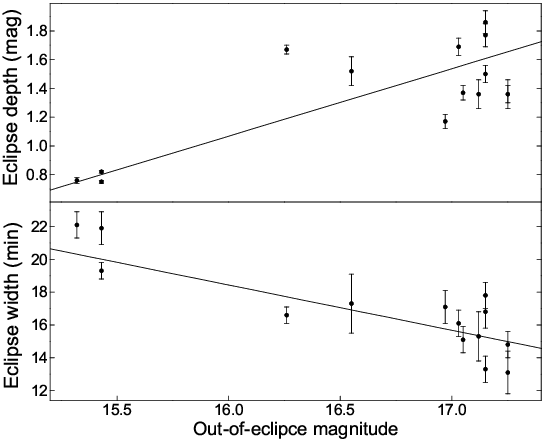}
\caption{Eclipse depth (top panel) and width (bottom panel) plotted against the out-of-eclipse magnitude of J0359}
\label{fig2}
\end{figure}

The photon noise in the differential light curves of J0359 varied depending on its brightness, which changed significantly during observations. When the out-of-eclipse magnitude of J0359 was around 15~mag, the photon noise varied between 0.04 and 0.05~mag outside of eclipses and between 0.08 and 0.09~mag during the deepest parts of the eclipses. As the out-of-eclipse magnitude of J0359 increased to about 17~mag, the photon noise increased by a factor of 4 to 9.

In Fig.~\ref{fig1} and other suitable places, I converted differential magnitudes into magnitudes using the $G$-band magnitude of the comparison star Gaia DR3 229958703719651200 \citep{gaia16, gaia23}. The $G$ band used in the $Gaia$ mission corresponds approximately to the wavelength range employed in my observations using photomultipliers, making it a suitable choice for this purpose. Additionally, the colour of the comparison star is similar to that of J0359, with only a difference of 0.21 mag in $BP-RP$, indicating minimal discrepancies between the photometric systems.

To make the detected eclipses more clear, Fig.~\ref{fig1} shows the light curves of J0359 with a time resolution of 128~s. The photon noise in these light curves is lower by a factor of $\sqrt{8}$ than that with a time resolution of 16~s. In addition to the 3 pairs of eclipses shown in Fig.~\ref{fig1}, 8 more eclipses were recorded in the shorter light curves (Table~\ref{tab2}). A total of 14 eclipses were recorded.

I determined the out-of-eclipse magnitudes by averaging segments of the light curves. They are included in Table~\ref{tab1}. To calculate the eclipse parameters, I used a time resolution of 64~s. If I had used a higher time resolution, it would have been impossible to calculate the magnitudes of the deepest points of eclipses due to large fluctuations in the number of recorded photons. The light curves with a time resolution of 64~s are available at the following link for further analysis: https://www.researchgate.net/publication/383988116. I measured the depth, width at half-depth and mid-eclipse time of the observed eclipses using Gaussian fits to eclipse profiles. The results of these measurements are presented in Table~\ref{tab2}.

As seen in Table~\ref{tab2}, the eclipse depths vary significantly ranging from 0.75$\pm$0.01 to 1.86$\pm$0.08~mag. The average depth is 1.35$\pm$0.10~mag. In the top panel of Fig.~\ref{fig2}, the depth is plotted against the out-of-eclipse magnitude of J0359. The first three points corresponding to the maximum brightness of J0359 significantly deviate from the other 11 points. These deviations exceed the error bars by many times. Therefore, it is no doubt that the depth of eclipses increases as the brightness of J0359 decreases. However, there is no correlation between the depth of eclipses and the brightness of J0359 for the remaining 11 data points. These data points show only a large scatter.

The eclipse widths at half-depth also show significant variations ranging from 13.1$\pm$1.3 to 22.1$\pm$0.8~min. The average width is 16.9$\pm$0.7~min. In the bottom panel of Fig.~\ref{fig2}, the width is plotted against the out-of-eclipse magnitude of J0359. This graph shows the correlation between the width of eclipses and the brightness of J0359. Eclipses become narrower as the brightness decreases. By assigning weights equal to the width error, a linear relationship was found between the width and out-of-eclipse magnitude: eclipse width~(min)~= $(63\pm5) - (2.8\pm0.3) \, \times$ out-of-eclipse magnitude. This relationship holds even without assigning weights, resulting in a slope of~$-3.2\pm0.5$.

It is believed that fitting an ephemeris to match eclipse times provides the highest precision for determining the period. To use this technique, I need a preliminary ephemeris. Successive eclipses occurring in pairs allow me to directly calculate the period using the mid-eclipse time of each eclipse. The average period calculated from three eclipse pairs (Fig.~\ref{fig1}) is 0.2284(6)~d, but this period is not precise enough. The accumulated error of this period is one oscillation cycle in 90~d. The validity time of the ephemeris is about four times shorter than the accumulated error (e.g., K. Mukai, https://asd.gsfc.nasa.gov/Koji.Mukai/iphome/iphome.html). It covers only 20~d. This time is much shorter than the observation time of 13 months. A more precise period is needed to derive a better preliminary ephemeris. 


\begin{figure}[t]
\centering
\includegraphics[width=84mm]{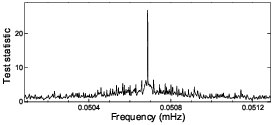}
\caption{Analysis of Variance spectrum of J0359}
\label{fig3}
\end{figure}
A much more precise period can be determined using the analysis of variance (AoV) method proposed by \citet{schwarzenberg89}. For non-sinusoidal signals, such as eclipses, the AoV spectrum is preferable to the Fourier power spectrum, in which the power is spread over many high-frequency harmonics \citep{schwarzenberg98}. Unfortunately, there is no simple method currently available to calculate period errors in AoV spectra. Therefore, I used the method proposed by \citet{schwarzenberg91} to calculate period errors in Fourier power spectra. However, in my previous studies, I found that this method overestimated the period error several times when applied to AoV spectra (e.g., \citealt{kozhevnikov21}).

\begin{figure}[t]
\centering
\includegraphics[width=84mm]{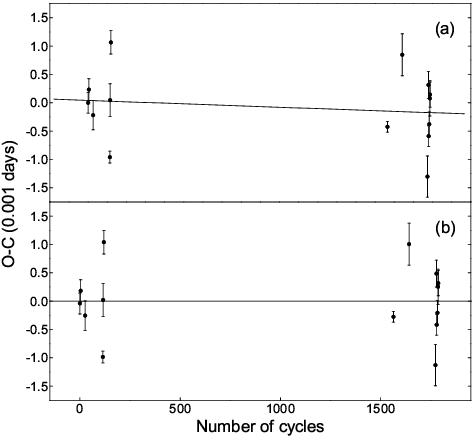}
\caption{(a) The (O--C) diagram for the preliminary ephemeris shows a noticeable slope. (b) The (O--C) diagram for the corrected ephemeris does not show any slope}
\label{fig4}
\end{figure}

\begin{table}[t]
\caption[ ]{Verification of the ephemeris}
\label{tab3}
\begin{tabular}{@{}l c c c}
\noalign{\smallskip}
\hline
\noalign{\smallskip}
Date (UT) & \scriptsize BJD$_{\rm TDB}$ mid-ecl.  & Number   &  O--C$\times 10^{-3}$ \\
                          & (-2459000)                               & of cycles  &        (d)          \\
\noalign{\smallskip}
\hline
\noalign{\smallskip}
2021 Oct. 7         & 495.30445     & 0       & -0.04$\pm$0.19   \\
2021 Oct. 8        & 496.21805      & 4       & 0.18$\pm$0.19    \\
2021 Oct. 13       & 501.24118     & 26      & -0.26$\pm$0.26  \\
2021 Nov. 2a      & 521.33470     & 114     & -0.99$\pm$0.10  \\
2021 Nov. 2b      & 521.56405     & 115     & 0.02$\pm$0.29    \\
2021 Nov. 3        & 522.47845     & 119     & 1.04$\pm$0.21   \\
2022 Sep. 29      & 852.43400     & 1564   & -0.28$\pm$0.10   \\
2022 Oct.17       & 870.24610     & 1642   & 1.01$\pm$0.37   \\
2022 Nov. 16       & 900.15701    & 1773   & -1.13$\pm$0.36   \\
2022 Nov. 17a     & 901.30034    & 1778   & 0.49$\pm$0.24   \\
2022 Nov. 17b     & 901.52778    & 1779   & -0.42$\pm$0.19  \\
2022 Nov. 18       & 902.44137    & 1783   & -0.21$\pm$0.22  \\
2022 Nov. 19a     & 903.12686    & 1786   & 0.25$\pm$0.31    \\
2022 Nov. 19b     & 903.35527    & 1787   & 0.32$\pm$0.22    \\
\noalign{\smallskip}
\hline
\end{tabular}
\end{table}

The AoV spectrum for all the data from J0359 is shown in Fig.~\ref{fig3}. I used a Gaussian fit to the peak to determine the maximum and error. The resulting period is 0.228\,343\,97(100)~d. This period has sufficient precision for a preliminary ephemeris. To obtain a preliminary ephemeris, I used the mid-eclipse time of the first eclipse. I then calculated the observed minus calculated (O--C) values for all mid-eclipse times (see Table~\ref{tab3}). The (O--C) diagram presented in Fig.~\ref{fig4}(a) shows that the points are quite scattered compared to their errors, which may be caused by the flickering of the star. Therefore, it is not appropriate to weigh (O--C) values based on errors. Instead I gave equal weight to each (O--C) value and obtained a linear relationship: $O-C = 0.000\,04(28) - 0.000\,000\,12(21) {\it E}$. As seen in Fig.~\ref{fig4}(a), this graph has a noticeable slope. After correcting for this slope and a small offset along the vertical axis, I obtained a final ephemeris with no slope or offset (Fig.~\ref{fig4}(b)). The final ephemeris is:

{\scriptsize
\begin{equation}
BJD_{\rm TDB}(\rm mid-ecl.) = 245\,9495.304\,49(28)+0.228\,343\,85(21) {\it E}. 
\label{ephemeris1}
\end{equation} }

The final orbital period, determined by the linear fit to the (O--C) values, is 0.228\,343\,85(21)~d. This period is close to that obtained from the AoV spectrum, but its error is five times less. The overestimation of the error in the AoV spectrum is consistent with my previous studies (e.g., \citealt{kozhevnikov21}). The accumulated error of the period reaches one oscillation cycle in 680~yr. Although the validity time of Ephemeris~\ref{ephemeris1} is four times shorter than this time,  nevertheless, it remains valid for a sufficiently long time up to 170~yr.

\begin{figure}[t]
\centering
\includegraphics[width=84mm]{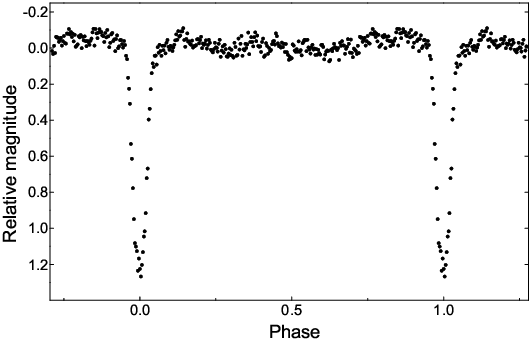}
\caption{Light curves folded with the orbital period of J0359}
\label{fig5}
\end{figure}

Figure~\ref{fig5} shows the light curves of J0359 folded with the orbital period. The time interval between points is 64~s. The errors for points outside the eclipse are about 0.04~mag, while the errors for points inside the eclipse are 0.11~mag. The eclipse depth in the folded light curve, as determined by a Gaussian fit, is 1.34$\pm$0.02 mag, which is consistent with the average eclipse depth obtained from individual eclipses presented in Table~\ref{tab2}.

\begin{figure}[t]
\centering
\includegraphics[width=84mm]{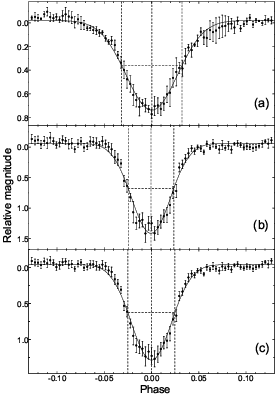}
\caption{Average eclipse profiles depending on the brightness of J0359: The average out-of-eclipse magnitude was 15.38$\pm$0.06~mag (a) and 17.10$\pm$0.04~mag (b). Frame (c) shows the average eclipse profile obtained by combining all observations. Solid lines represent the fitted Gaussian curves, while dashed lines indicate the width of the eclipses at half-depth}
\label{fig6}
\end{figure}

As shown in Fig.~\ref{fig2}, the depth and width of eclipses vary depending on the brightness of J0359. To investigate how eclipse profiles change, I have identified two groups of light curves: those when J0359 is near its maximum brightness (the bright state) and those when it is near its minimum brightness (the faint state). The first group contained 2 light curves with out-of-eclipse magnitudes of 15.32$\pm$0.02 and 15.43$\pm$0.04~mag, with an average magnitude of 15.38$\pm$0.06~mag. There were 3 eclipses in the first group. The second group consisted of 7 light curves with out-of-eclipse magnitudes ranging from 16.97$\pm$0.04 to 17.25$\pm$0.08~mag, with an average magnitude of 17.10$\pm$0.04~mag. There were 9 eclipses within the second group. I obtained average eclipse profiles by folding the light curves. When J0359 was in the bright state, the profile had a V-shape with long wings and a sharp bottom (see Fig.~\ref{fig6}(a)). As J0359 became dimmer, the profile changed to a U-shape, with shorter wings and a flatter bottom (see Fig.~\ref{fig6}(b)). The average eclipse profile obtained from all the light curves (see Fig.~\ref{fig6}(c)) resembled the profile when J0359 was in the faint state. Comparing the eclipse profiles with Gaussian fits, we can clearly see the differences. Using these fits, I calculated the eclipse widths at half-depth, which are shown by dashed lines in Fig.~\ref{fig6}. The widths are 0.0640(12), 0.0481(11) and 0.0497(9) phases in cases a, b and c, respectively. The eclipse depths are 0.76$\pm$0.02, 1.49$\pm$0.03 and 1.34$\pm$0.02~mag in cases a, b and c, respectively.

To search for coherent short-period oscillations similar to those observed in intermediate polars, I analysed the five longest light curves of J0359, which lasted longer than 7.4~h (see Table~\ref{tab1}). Eclipses were removed from these light curves by interpolating the magnitudes between the start and end of each eclipse. Overnight averages and variations with periods longer than the length of the light curve were excluded by subtracting a third-order polynomial fit. Fourier power spectra were calculated for these modified light curves.

\begin{figure}[t]
\centering
\includegraphics[width=84mm]{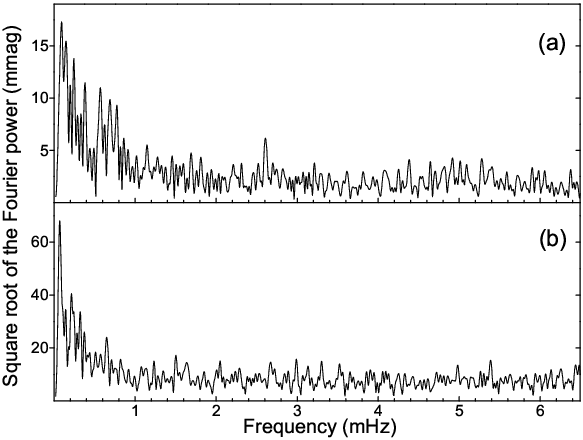}
\caption{Averaged power spectra of the longest light curves obtained during the bright (a) and faint (b) states of J0359}
\label{fig7}
\end{figure}

Because the light curves were collected at different brightness states of J0395, the calculated power spectra contain different levels of photon noise. Therefore, I divided the power spectra into two groups based on the brightness of J0359. The first group consisted of two power spectra calculated when J0359 was in the bright state, with an average out-of-eclipse magnitude of 15.38$\pm$0.06~mag. The second group consisted of three power spectra calculated when J0359 was in the faint state, with an average out-of-eclipse magnitude of 17.14$\pm$0.07~mag. The low-frequency parts of the averaged power spectra are shown in Fig.~\ref{fig7}. Neither power spectrum shows any significant peaks up to 31~mHz, indicating that there are no coherent short-period oscillations. However, both spectra do show red noise at low frequencies up to about 2~mHz likely due to orbital variations and flickering. The white noise levels at higher frequencies differ by a factor of approximately 4 between the two groups. This is consistent with the differences in the brightness of J0359.

\begin{figure}[t]
\centering
\includegraphics[width=84mm]{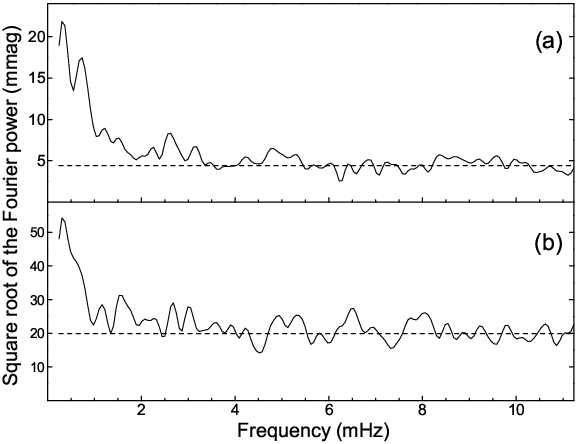}
\caption{Averaged power spectra of short segments of the light curves obtained during the bright (a) and faint (b) states of J0359. The dashed lines indicate the white noise level, which was calculated by averaging the power spectrum over a frequency range of 7 to 31~mHz}
\label{fig8}
\end{figure}

To improve the detection of red noise in the power spectra and isolate its high-frequency component \\ caused by flickering, I extracted short segments without eclipses from the five longest light curves. These segments had 256 points and lasted about 1.1~h. I removed any changes on the hourly time scale using a linear fit to eliminate low-frequency variations that were not caused by flickering. In total, I analysed 12 segments when J0359 was in the bright state and 17 segments when it was in the faint state. Figure~\ref{fig8} shows the low-frequency parts of the averaged power spectra of these segments in the two brightness states of J0359. As seen, the red noise is present up to frequencies of 6~mHz in the bright state of J0359 and 4~mHz in the faint state. 

To quantify the flickering seen in the light curves of J0359 using the red noise present in the power spectra, I applied Parseval's theorem. This allowed me to relate the noise in the time domain to the noise in the frequency domain, as described in \cite{kozhevnikoviz00}. The total variance of the light curve can be considered as the sum of variances caused by white noise and red noise: $\sigma^2_{\rm total} = \sigma^2_{\rm white} + \sigma^2_{\rm red}$. Using the Fourier power spectrum for each light curve segment, I calculated the total, white, and red noise. The average total noise was 59$\pm$3 and 233$\pm$3~mmag in the bright and faint states of J0359, respectively. These values are consistent with those obtained directly from the light curves. The average white noise was 49$\pm$3 and 216$\pm$15~mmag in the bright and faint statistics, respectively. These values are close to the photon noises calculated from counts. The average red noise was 33$\pm$2 and 82$\pm$6~mmag in the bright and faint states, respectively.

Although red noise may be caused by atmospheric effects, the atmospheric red noise in our three-channel photometer is low, not exceeding 2.7~mmag (see table 1 in \citealt{kozhevnikoviz00}). Therefore, I conclude that the rapid changes in the light curves of J0359 (Fig.~\ref{fig1}) are most likely due to the flickering of the star itself and not to atmospheric effects. The photon noise was much higher during the faint state of J0359, masking the flickering. However, the noise analysis showed that the flickering power increased noticeably as J0359 dimmed. This pattern was not obvious in the light curves presented in Fig.~\ref{fig1}.

\section{Discussion}\label{sec4}


J0359 has been identified as a candidate CV based on its emission-line spectrum as reported by \citet{hou20}. However, without analysing the photometric properties of this object, it is not possible to confirm whether it is a true CV. During my photometric observations, I detected eclipses, which indicated that J0359 is a close binary system with a short orbital period, typical of CVs. The orbital period was measured to be 0.228\,343\,85(21)~d. The eclipse profile has extended wings, also typical of CVs. The depth and the width of the eclipses vary with the brightness of J0359, suggesting the presence of a disc. Based on these findings, I conclude that J0359 is definitively a CV, and further confirmation is provided by the observed flickering.

The variability of the eclipse profile indicates that the size of the disc is changing. As J0359 becomes dimmer, the bottoms of the eclipses seem to flatten, indicating that the eclipses may be close to a total eclipse (Fig.~\ref{fig6}(b)). This can be confirmed by comparing the expected light emitted by the red dwarf with the total light during eclipses. Using equation 2.102 and Fig.~2.46 in \citep{warner95}, I calculated $M_{\rm V}$ of the red dwarf to be 8.5$\pm$0.3~mag. 

To convert $M_{\rm V}$ into $M_{\rm G}$, I selected 200 stars from the Hipparcos catalogue (\citealt{ESA97}, https://vizier.cds.unistra.fr/viz-bin/VizieR-3?-source=I/239/) that were fainter than 9~mag and were randomly distributed in declination. I excluded variables, binaries, and stars without $B-V$ colour measurements. Using the remaining 155 stars and their photometric data from the Gaia DR3 catalogue (\citealt{gaia16, gaia23}, https://vizier.cds.unistra.fr/viz-bin/VizieR-3?-source=I/355/gaiadr3), I derived a relationship between the $V$- and $G$-band magnitudes: $V - G = 0.015 + 0.180 (B - V) - 0.0050 (B-V)^2 + 0.10862  (B-V)^3$ with a typical scatter of $\pm$0.05~mag for $B-V < 1.1$~mag  and $\pm$0.1~mag for $B-V > 1.1$~mag.

As seen in Fig~3 in \citet{perryman97}, a red dwarf with $M_{\rm V}$ = 8.5~mag has a $B-V$ colour of 1.40$\pm$0.10~mag. However, according to the Gaia DR3 catalogue, the interstellar reddening for J0359, $E(BP-RP)$, is significant and estimated to be 0.62$\pm$0.03~mag. Because the effective wavelength difference between $BP$ and $RP$ is about twice that between $B$ and $V$ (see Fig.~1 in \citealt{ritter20}), it is reasonable to increase the $B-V$ colour by half the value of $E(BP-RP)$. I corrected the $B-V$ colour of the white dwarf to 1.70$\pm$0.10~mag and converted $M_{\rm V}$ into $M_{\rm G}$, using the above $V-G$ relationship, $M_{\rm G}$=7.66$\pm$0.36~mag. Using a distance of 1004$\pm$44~pc and an interstellar extinction of 1.10$\pm$0.06~mag from the Gaia DR3 catalogue, I calculated the apparent $G$-band magnitude of the red dwarf to be 18.77$\pm$0.37~mag. When J0359 was in the faint state, the magnitude of the deepest part of the average eclipse was 18.59$\pm$0.09~mag (Fig.~\ref{fig6}(b)). The difference between these two magnitudes was 0.18$\pm$0.38~mag, and the light coming from the disc was negligible, indicating that the eclipse shown in Fig.~\ref{fig6}(b) was nearly total.

\begin{figure}[t]
\centering
\includegraphics[width=84mm]{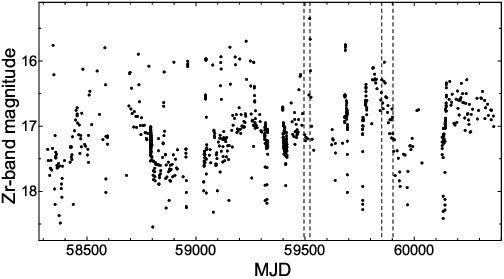}
\caption{Long-term light curve of J0359 based on data from the ZTF survey. Dashed lines indicate the times of my observations}
\label{fig9}
\end{figure}

The eclipse profile in J0359 changed from a V-shape to a U-shape as the brightness decreased (see Fig.~\ref{fig6}). This is similar to what was observed in dwarf novae during their outbursts (see Fig.~3 in \citealt{baptista00}). However, during my observations, the brightness of J0359 outside of eclipses varied between 15.32$\pm$0.04 and 17.25$\pm$0.08~mag, which is lower than typical dwarf nova outburst amplitudes. Fortunately, the Zwicky Transient Facility (ZTF) survey (https://irsa.ipac.caltech.edu/Missions/ztf.html) has a long-term light curve of J0359 that can help us to understand its nature. This light curve shown in Fig.~\ref{fig9} spans 5.6 yr and shows 28 dips to 18--18.5~mag, which are identified as eclipses according to Ephemeris~\ref{ephemeris1}. The left part of the light curve shows 18 outbursts, each with an amplitude of 1--1.5~mag and an average time interval between them of 50$\pm$10~d. Based on the Kukarkin-Parenago relationship for dwarf novae, the expected amplitude of outbursts occurring at such time intervals is around 4~mag \citep{warner95}. However, the observed amplitudes are significantly lower, resembling those of stunted outbursts in nova-like variables, which can reach up to 1~mag in amplitude \citep{honeycutt01}. Surprisingly, there have been no significant outbursts visible in the light curve over the past 1.6~yr since MJD~59760.

To determine whether J0359 is a dwarf nova or a nova-like variable, I calculated the time-averaged absolute magnitude of its disc. I used the epoch photometry obtained by the $Gaia$ mission (https://vizier.cds.unistra.fr/viz-bin/VizieR-3?-source=I/355/epphot). The $G$-band light curve of J0359 spans 2.5~yr and shows three small outbursts. The average $G$-band magnitude is 17.04$\pm$0.09~mag, and the average $BP-RP$ colour is 1.30$\pm$0.05~mag. I cannot exclude eclipses directly from the light curve as they were smeared out during data collection. However, I can use the folded light curve in Fig.~\ref{fig5} to see what happens when I exclude them. After excluding them, the average magnitude decreased by 0.06~mag, resulting in a corrected average magnitude of 16.98$\pm$0.09~mag. Using the distance and interstellar extinction mentioned above, I calculated the average absolute $G$-band magnitude to be 5.87$\pm$0.14~mag. To convert the $G$-band magnitude into the $V$-band magnitude using the aforementioned 155 stars from the Hipparcos catalogue, I derived another relationship between the $V$- and $G$-band magnitudes: $V - G = 0.015 + 0.052(BP - RP) + 0.1463(BP - RP)^2 + 0.00773(BP - RP)^3$ with a typical scatter of $\pm$0.05~mag. Using this relationship, the average absolute $V$-band magnitude of J0359 is 6.22$\pm$0.15~mag. After removing the light from the red dwarf, I calculated the average absolute $V$-band magnitude of its disc to be 6.36$\pm$0.15~mag.

The absolute magnitude of the disc should be corrected for the inclination of the orbit. Using the ratio of the masses of the red and white dwarfs and the duration of the eclipse of the white dwarf, it is possible to estimate the orbital inclination using Fig.~2 in \citet{horne85}. Even if the exact duration of the eclipse of the white dwarf is not known, the eclipse width at half-depth can still provide a good estimate (\citealt{warner95}, Chapter 2.6.2). The maximum mass ratio for stable mass transfer is 2/3 \citep{warner95}. To find the minimum mass ratio, I assume that the mass of the white dwarf is 1.44$M_{\rm Sun}$  (Chandrasekhar limit) and use the orbital period of J0359 and equation 2.100 in \citet{warner95} to calculate the mass of the red dwarf, yielding a value of $0.54\pm0.02 M_{\rm Sun}$ and a minimum mass ratio of 0.38. When J0359 was in the faint state, the width of the eclipse at half-depth was 0.0481(11) phases (Fig.~\ref{fig6}(b)). For this width, Fig.~2 in \citet{horne85} provides a range of possible orbital inclinations between $72.8^\circ$ and $76.0^\circ$. Using equation 2.63 in \citet{warner95}, this range of inclinations corresponds to a range of disc corrections between 0.92 and 1.20~mag. After applying these corrections, the average absolute $V$-band magnitude of the disc, corrected for standard inclination, ranges from 5.16$\pm$0.15 to 5.44$\pm$0.15~mag.

In Fig.~3.9 in \citet{warner95}, the solid curve shows the time-averaged absolute $V$-band magnitude of discs corresponding to the critical mass transfer rate. This is the rate below which discs are expected to be thermally unstable leading to dwarf nova outbursts. This curve intersects with a vertical line corresponding to an orbital period of 5.5~h at 5.2~mag. The range of average absolute $V$-band magnitudes of the disc in J0359 lies almost completely below this curve. Although J0359 exhibits low-amplitude outbursts, it should still be classified as a dwarf nova due to its relatively low average disc brightness. The low amplitudes of the outbursts can be explained by the fact that the mass transfer rate is close to the limit of thermal instability. This is similar to the case of the dwarf nova IX~Vel, where the mass transfer rate is close to the limit of thermal instability. As a result, only the outer part of the disc becomes thermally unstable, causing low-amplitude outbursts \citep{kato21}.

As seen in Fig.~\ref{fig9}, there have been no significant outbursts in the light curve over the past 1.6~yr since MJD~59770. During this time, the average brightness was 0.3~mag brighter than in the left part of the light curve where the outbursts were clearly visible. It appears that in this part of the light curve, the mass transfer rate slightly exceeds the limit of thermal instability. Therefore, we can expect behaviour similar to that of Z~Cam stars, which have standstill periods when the mas transfer rate during a standstill only slightly exceeds its long-term average \citep{warner95}. However, as seen in Fig.~\ref{fig9}, there is a gradual decrease in brightness from 16.2 to 17.9~mag, rather than a standstill. This decrease lasts for 120~d between MJD~59810 and MJD~59930. A decrease in brightness, such as this, is typical of nova-like variables of the VY~Scl type. It appears that J0359 may be a dwarf nova at times and a VY~Scl star at other times.

\section{Conclusion}\label{sec5}

I conducted photometric observations of the candidate cataclysmic variable J0359 and discovered previously unknown eclipses. These eclipses, along with other photometric characteristics, confirmed that it is indeed a cataclysmic variable. However, it has some unusual properties. After carefully analysing my photometric data collected over 11 nights spanning 13 months, I found the following:

\begin{enumerate}

\item Due to the extensive coverage of observations and the significant number of recorded eclipses, I precisely determined the orbital period, $P_{\rm orb}=0.228\,343\,85\pm0.000\,000\,21$~d.

\item For the eclipses, I derived an ephemeris. This ephemeris has a long validity time and is suitable for studying changes in the orbital period. It can also be useful for measuring radial velocity in future spectroscopic observations.

\item The eclipses showed changes in their depth and width, which were related to variations in the brightness of J0359. As the brightness decreased, the eclipses became deeper and narrower. The eclipse depth varied between 0.75$\pm$0.01 and 1.86$\pm$0.08~mag, and the eclipse width at half-depth varied between 13.1$\pm$1.3 and 22.1$\pm$0.8~min.

\item During my observations, the brightness of J0359 outside eclipses varied between 15.32$\pm$0.02 and 17.25$\pm$0.08~mag and did not show large changes typical of dwarf novae. Based on the light curve from the ZTF survey, J0359 showed only small outbursts with amplitudes of 1--1.5~mag, which are significantly lower than typical amplitudes of dwarf nova outbursts.

\item Using the average eclipse width at half-depth, I estimated the possible range of orbital inclinations to be $72.8^\circ$--$76.0^\circ$. The range of absolute $V$-band magnitudes of the disc was found to be between 5.16$\pm$0.15 and 5.44$\pm$0.15~mag. For a CV with an orbital period of 5.5~h, the average absolute $V$-band magnitude of the disc at which the disc becomes thermally unstable is 5.2~mag. Because the average disc brightness in J0359 is fainter than this, this CV should be classified as a dwarf nova. The low amplitudes of the outbursts can be explained by the fact that the mass transfer rate is close to the limit of thermal instability.

\item As seen in the light curve of J0359 from the ZTF survey, there have been no significant outbursts over the past 1.6~yr. During this time, J0359 stayed on average 0.3~mag brighter than during times when outbursts were visible. This suggests that the average disc brightness slightly exceeded the limit of thermal instability. However, instead of the expected standstill seen in Z~Cam stars, J0359 showed a gradual decrease in brightness over 120~d, which is typical of nova-like variables of the VY~Scl type. It appears that J0359 may be a dwarf nova at times and a VY~Scl star at other times.

\end{enumerate}

\section*{Acknowledgments}

The work of V. P.~Kozhevnikov was supported by the Ministry of science and higher education of the Russian Federation, agreement FEUZ-2023-0019. The author is grateful to the ZTF team for making their data available to the public. This work has made use of NASA's Astrophysics Data System Bibliographic Services, the VizeR catalogue access tool \citep{ochsenbein00} and the Aladin sky atlas developed at CDS, Strasbourg observatory, France \citep{bonnarel00, boch14}. This work has made use of data from the European Space Agency (ESA) mission $Gaia$ (https://www.cosmos.esa.int/gaia), processed by the $Gaia$ Data Processing and Analysis Consortium (DPAC, https://www.cosmos.esa.int/web/gaia/dpac/consortium). Funding for the DPAC has been provided by national institutions, in particular the institutions participating in the $Gaia$ Multilateral Agreement.

\subsection*{Author contributions}
I, Valerij P. Kozhevnikov, is the only author of the manuscript. All content of the manuscript is prepared by me.
\subsection*{Financial disclosure}

None reported.

\subsection*{Conflict of interest}

The authors declare no potential conflict of interests.

\bibliography{kozhevnikov_ms}%



\end{document}